\documentclass[aps,prl,reprint,showkeys,nofootinbib]{revtex4-2}

\usepackage{amsmath,amssymb,bm}
\usepackage{booktabs}
\usepackage{graphicx}
\usepackage{xcolor}
\usepackage{tikz}
\usetikzlibrary{arrows.meta,positioning,calc}
\usepackage[colorlinks=true,citecolor=blue,linkcolor=blue,urlcolor=blue]{hyperref}

\definecolor{dblue}{RGB}{34,92,153}
\definecolor{pred}{RGB}{184,58,63}
\definecolor{softgray}{RGB}{245,246,248}

\usepackage{subfigure}

\begin{document}

\title{Floquet engineering of competing antiferromagnetism and\\
$d$-wave superconductivity on the square lattice}

\author{Zhaoyu Han}
\affiliation{Department of Physics, Harvard University, Cambridge MA 02138, USA}

\author{Subir Sachdev}
\affiliation{Department of Physics, Harvard University, Cambridge MA 02138, USA}

\begin{abstract}
We propose a driven Lieb--Hubbard quantum simulator whose prethermal dynamics realize a square-lattice spin-1/2 fermion model with independently tunable repulsive on-site and attractive bond interactions. Periodically modulating the charge-transfer offset between the site ($d$) and bond ($p$) orbitals of the Lieb lattice brings a $p$-orbital doublon energetically close to a pair in the $d$ manifold while keeping all $p$-orbital singlons off resonance, thereby creating a synthetic ``negative-$U$'' center on a Lieb-lattice bond site. Two controlled eliminations then generate a compact bond attraction in the reduced $d$-only model on the square lattice, despite the microscopic repulsion in the $p$ orbital. The resulting interaction $J$ is tunable independently of the Hubbard repulsion $U_d$ on the $d$ orbitals, while interference between photon-assisted paths provides access to an intermediate-coupling regime in which $J$ and $U_d$ are both comparable to the effective hopping. At half filling, a mean-field calculation in this regime finds adjacent antiferromagnetic and $d$-wave superconducting phases, as well as narrow coexistence regions, suggesting close competition between these orders. We discuss the branch-preparation, prethermal, and higher-band conditions required to translate the formal construction into an optical-lattice protocol. More broadly, our work identifies a structural similarity between Floquet systems and electron–phonon problems that may guide the design of novel quantum-simulation protocols.
\end{abstract}

\maketitle

\textit{Introduction.---} Ultracold fermions in optical lattices provide a highly controlled setting for quantum simulations of strongly correlated electronic systems. Rapid progress in this field~\cite{Esslinger2010,GrossBloch2017,Schaefer2020,Hart2015,Mazurenko2017}, including significantly lowered effective temperatures and access to doped frustrated magnets~\cite{Xu2025}, raises the prospect of studying the competition among orders relevant to the cuprates, such as $d$-wave superconductivity ($d$SC), antiferromagnetism (AF), and density-wave order. Their interplay remains a central puzzle in condensed-matter physics. The Hubbard model is a canonical arena for these phenomena~\cite{HubbardReview} and has been extensively studied in optical-lattice platforms~\cite{Esslinger2010,GrossBloch2017,Schaefer2020}. The status and scale of its putative $d$SC phase nevertheless remain numerically delicate: the most controlled evidence is clearest at weak coupling, while the intermediate-coupling ground state remains under active debate~\cite{Qin2020,VilardiBonettiMetzner2020,Roth2025}. This motivates a complementary perspective, particularly timely as quantum simulators acquire increasingly flexible controls, including Floquet driving~\cite{Eckardt2017,Messer2018,RubioAbadal2020,Tsuji2011,Mentink2015,Takahashi2025}: can one engineer a structured interaction that exposes the competition between AF and $d$SC more directly?

This goal is not merely materials-specific; it also addresses fundamental questions in quantum many-body physics. The AF--$d$SC boundary is a candidate setting for deconfined quantum criticality (DQCP)~\cite{VojtaZhangSachdev2000,Christos2023,Christos2024}, whose existence and critical behavior remain under intense investigation. Recent sign-problem-free simulations of fermionic spinons and bosonic chargons coupled to a fluctuating SU(2) gauge field have found strong evidence for such a transition~\cite{ChenSachdevMeng2026} at half filling.

Recent studies on electron--phonon models provide an important design lesson.  Integrating out a fast phonon that modulates a lattice bond produces the square of a bond electronic operator in the effective electron-only model on the vertex lattice. Its decomposition can contain AF exchange, correlated hopping, and pair hopping, so the mediated coupling can promote both magnetic order and mobile Cooper pairing. This structure underlies the AF--$d$SC competition in a bond-interaction model~\cite{Assaad1998}, in Su--Schrieffer--Heeger--Hubbard models~\cite{WangJiangYao2025,CaiLiYao2021,GoetzAssaad2022,YangWang2022}, and in a Lieb--Holstein model~\cite{HanKivelson2023}. The same lesson appears in more microscopic studies of cuprates: a $B_{1g}$ buckling phonon can enhance $d$-wave pairing when AF and pairing tendencies are already close~\cite{HonerkampFuLee2007}, consistent with the strongly anisotropic oxygen-mode vertices emphasized in Refs.~\cite{Devereaux2004,Johnston2010}. A real-space interpretation of these results is Anderson's negative-$U$-center mechanism~\cite{Anderson1975}: a high-energy orbital that favors double occupation can lower the energy of an itinerant pair through virtual pair transfer even when that orbital is absent from the low-energy manifold. In the electron--phonon realizations, lattice vibration creates such attractive centers on bonds, and then a modest repulsive Hubbard interaction can suppress the competing on-site $s$-wave channel and expose the bond-generated $d$SC and AF tendencies. Recently, studies on similar e-ph models also found resonating-valence-bond (RVB) phases with emergent gauge structure~\cite{HanKivelson2023,doi:10.1073/pnas.2426111122,doi:10.1073/pnas.2421778122}, suggesting an emergent gauge theory description underlying the physics of such systems.

\begin{figure}[t!]
\centering

\subfigure[]{\includegraphics[width=0.4\linewidth]{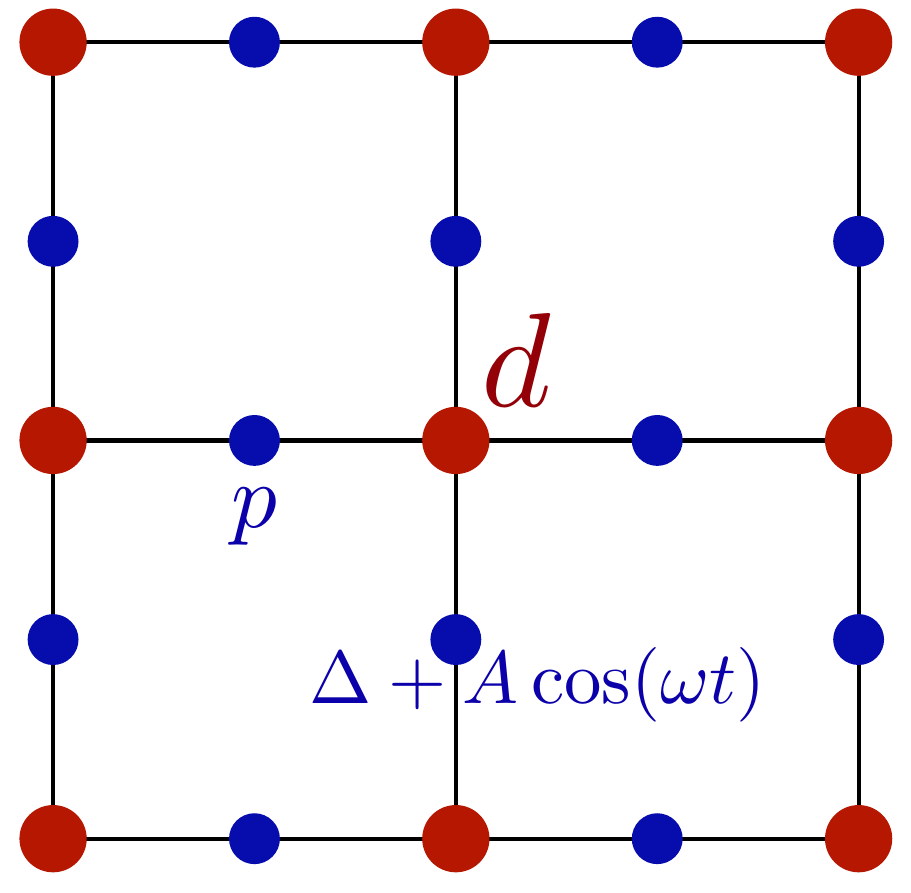}}
\\
\subfigure[]{
\resizebox{0.8\linewidth}{!}{%
\begin{tikzpicture}[
  level/.style={line width=1.1pt},
  darr/.style={-{Latex[length=2.2mm]},line width=1.1pt,dblue},
  rarr/.style={-{Latex[length=2.2mm]},line width=1.1pt,pred},
  outcome/.style={draw=black!55,rounded corners=2pt,fill=softgray,
    align=center,inner sep=5pt},
  every node/.style={font=\scriptsize}
]
\begin{scope}[xshift=0cm]
  \draw[level,black!75] (0.35,0) -- (3.05,0)
    node[right,black,align=left] {site-manifold};
  \draw[level,dblue,dashed] (0.35,1.35) -- (3.05,1.35)
    node[right,black,align=left] {bond singlon};
  \draw[level,pred,dashed] (0.35,3.2) -- (3.05,3.2)
    node[right,black,align=left] {bond doublon};
  \draw[<->,black!45] (0.65,0.03) -- (0.65,1.32)
    node[midway,left,yshift = -5] {$\Delta$};
  \draw[<->,black!45] (1.,0.03) -- (1.,3.17)
    node[midway,right,yshift = 20] {$2\Delta+U_p$};
\draw[darr,pred] (2.8,3.15) -- (2.8,0.85)
    node[midway,right,align=left] {Sambe folding\\$-\hbar\omega$};
\draw[level,pred,dashed] (0.35,0.8) -- (3.05,0.8)
    node[right,black,align=left] {photon-dressed\\bond doublon};
\draw[->, thick, bend left=40,pred, densely dashed] (2.64, 0.05) to (2.64, 0.75);
\draw[->, thick, pred, densely dashed] (2.66, 0.75) to[bend left=40] node[midway, right,align=left] {$J$} (2.66, 0.05);

\draw[->, thick, bend left=25,dblue, densely dashed] (1.64, 0.05) to (1.64, 1.3);
\draw[->, thick, dblue, densely dashed] (1.68, 1.3) to[bend left=25] node[midway, right,align=left] {$t_\text{eff}$} (1.68, 0.05);
\end{scope}

\end{tikzpicture}
}%
}

\caption{(a) Lieb lattice with $d$ orbitals on the vertices of a square lattice and $p$ orbitals at the bond centers.
Without driving, singly and doubly occupied $p$ configurations are both high in energy, and all microscopic interactions are repulsive.
(b) Illustration of the mechanism. The site manifold consists of the square lattice of $d$ orbitals. Modulation of the charge-transfer gap folds the one-photon $p$-doublon replica to a small positive energy $\delta_R$ above the $d$-bond pair, while single-$p$ processes remain off resonance. Eliminating this Floquet-created closed channel yields a compact bond attraction whose operator content includes AF exchange and pair hopping. The negative-$U$-center analogy is therefore dynamical and branch resolved.}
\label{fig:scheme}
\end{figure}
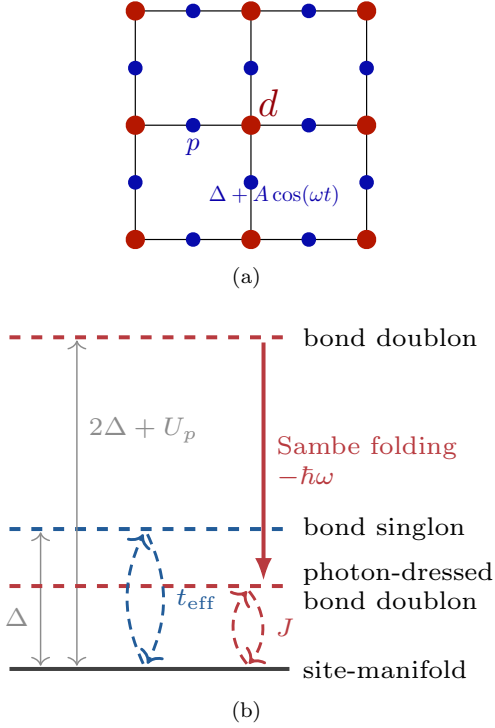

Our proposal asks whether a coherent Floquet drive can generate a similarly structured effective interaction. We begin with a Lieb--Hubbard model, a lattice geometry that has recently been realized in optical-lattice platforms~\cite{Taie2015,Lebrat2026}. The construction is a driven negative-$U$-center mechanism built from a nearly resonant bond-orbital doublon. Modulating the charge-transfer gap between the site ($d$) and bond ($p$) orbitals uses quasienergy periodicity to bring a one-photon replica of the $p$ doublon close to the $d$-orbital branch, while the corresponding $p$-singlon states remain off resonance [Fig.~\ref{fig:scheme}]. For positive dressed detuning, virtual occupation of this synthetic center lowers the energy of a $d$-orbital pair and produces an interaction reminiscent of those generated by electron--phonon coupling on bonds. A mean-field calculation at half filling shows that the resulting $d$-only model supports both $d$SC and AF in its intermediate-coupling phase diagram.

This setup receives additional motivation from light-induced superconducting phenomena in the cuprates (or more generally phonon driven floquet matters~\cite{10.1021/acs.nanolett.7b05391}). Photoemission reveals a widespread $50$--$80$ meV kink consistent with coupling to oxygen vibrations~\cite{Lanzara2001}, while symmetry-resolved theory and experiment distinguish a predominantly antinodal $B_{1g}$ buckling coupling from a more nodal bond-stretching coupling~\cite{Devereaux2004,Johnston2010,Reznik2006}. Resonant mid-infrared excitation of lattice modes in stripe-ordered and underdoped cuprates produces transient superconducting-like interlayer optical responses~\cite{Fausti2011,Hu2014}, and time-resolved diffraction connects these responses to driven apical-oxygen and bond-buckling distortions~\cite{Mankowsky2014}. If bond-buckling modes locally modulate the charge-transfer gap, the present cold-atom construction may provide a useful caricature of this mechanism.

\textit{The model.---} We consider a Lieb--Hubbard model with a periodically driven charge-transfer gap,
\begin{align}
H(\tau)=&
-t\sum_{\langle ij\rangle,\sigma}
\left[p^\dagger_{\langle ij\rangle,\sigma}
(d_{i\sigma}+d_{j\sigma})+\mathrm{h.c.}\right]
\nonumber\\
&+U_p\sum_{\langle ij\rangle}
n^p_{\langle ij\rangle,\uparrow}n^p_{\langle ij\rangle,\downarrow}
+U_d\sum_i n^d_{i\uparrow}n^d_{i\downarrow}
\nonumber\\
&+[\Delta+A\cos(\omega\tau)]
\sum_{\langle ij\rangle}n^p_{\langle ij\rangle}.
\label{eq:model}
\end{align}
Here $d_{i\sigma}$ annihilates a fermion on a square-lattice vertex and $p_{\langle ij\rangle,\sigma}$ annihilates one at a nearest-neighbor bond center. We take both $U_d$ and $U_p$ to be repulsive, as is natural for two hyperfine states with a positive scattering length. The reference filling is one fermion per unit cell. We assume
\begin{equation}
t,\ U_d\ll\Delta,U_p,\qquad
\hbar\omega\simeq 2\Delta+U_p ,
\label{eq:regime}
\end{equation}
so that all one-particle and spatially separated two-particle $p$ excitations remain off resonance. The drive amplitude must also remain below the gap to unwanted Wannier bands. These assumptions are stated quantitatively in the End
Matter.

\textit{Floquet--boson correspondence.---}
The analogy between the Floquet problem and an electron--boson problem, such as an electron--phonon model, becomes explicit in the rotating frame. With
$\hat n_p=\sum_{\langle ij\rangle}n^p_{\langle ij\rangle}$ and
$\kappa=A/(\hbar\omega)$, the transformation
$W(\tau)=\exp[-i\kappa\sin(\omega\tau)\hat n_p]$ removes the modulation and
changes the hybridization to
\begin{equation}
-t\sum_{\langle ij\rangle,\sigma}
\left[e^{i\kappa\sin(\omega\tau)}
p^\dagger_{\langle ij\rangle,\sigma}(d_{i\sigma}+d_{j\sigma})
+\mathrm{h.c.}\right].
\label{eq:rotating}
\end{equation}
The phase has the Jacobi--Anger expansion
\begin{equation}
 e^{i\kappa\sin\omega\tau}
 =\sum_{m=-\infty}^{\infty}{\cal J}_m(\kappa)e^{im\omega\tau},
\label{eq:bessel}
\end{equation}
where ${\cal J}_m$ is the Bessel function of the first kind. In Holstein-type electron-phonon models, a Lang--Firsov transformation similarly dresses electron hopping with a phonon displacement operator, whose matrix elements generate Franck--Condon sidebands and integer-shifted virtual denominators~\cite{LangFirsov1962,HanKivelsonYao2020,HanKivelson2023}. In Sambe space, $m$ labels a synthetic ladder of photon replicas, and the Bessel amplitudes are the coherent-drive analogues of Franck--Condon weights. Unlike a zero-temperature phonon vacuum, however, a classical drive supplies and absorbs quanta symmetrically. It can therefore place a selected composite excitation at a small, sign-tunable quasienergy detuning. The drive thus creates a branch-selective resonant channel rather than simply reproducing a phonon-mediated attraction. We emphasize, however, that a classical drive generates neither a phonon cloud nor the associated polaronic dressing. In particular, interference among drive-assisted virtual processes permits the magnitudes, signs, and relative weights of the mediated hopping and interaction terms to be tuned more flexibly than in conventional electron–phonon settings.

\textit{The perturbation theory.---} Under the assumed hierarchy of scales, high-energy states can be eliminated through two successive Schrieffer--Wolff transformations. The first step removes singly occupied $p$ orbitals while retaining the nearly resonant $p$ doublon. For convenience, we define the bonding orbital of two neighboring sites $d_{[i+j]} \equiv (d_i+d_j)/\sqrt{2}$ and
\begin{align}
D^\dagger_{[i+j]}&=d^\dagger_{[i+j]\uparrow}d^\dagger_{[i+j]\downarrow},\\
P^\dagger_{\langle ij\rangle}&=p^\dagger_{\langle ij\rangle,\uparrow}
p^\dagger_{\langle ij\rangle,\downarrow}.
\end{align}
Here $D^\dagger_{[i+j]}$ creates a pair in the $d$ bonding orbital; we call it a \emph{$d$-bond pair}. $P^\dagger_{\langle ij\rangle}$ creates two fermions on the single bond-center $p$ orbital; we call this hard-core boson a \emph{$P$ dimer}. Second-order virtual processes produce nearest-neighbor hopping within the $d$ manifold, coherent  $D$-$P$ pair conversion and Lamb shift of detuning: 
\begin{align}
t_{\rm eff}
&=\frac{t^2}{\Delta}{\cal T} \ , \quad \quad \Lambda
=\frac{2t^2}{\Delta}{\cal L}, \\
\delta_R &= 2\Delta+U_p-\hbar\omega+\frac{4t^2}{\Delta}({\cal R}+{\cal T}) 
\label{eq:leading}
\end{align}
where we define dimensionless coefficients
\begin{align}
{\cal T}
&=\sum_m
\frac{{\cal J}_m(\kappa)^2}
{1-m\hbar\omega/\Delta}, \\
{\cal L}
&=\sum_m \frac{{\cal J}_m(\kappa){\cal J}_{1-m}(\kappa) [ -U_p/\Delta+(1-2m)\hbar\omega/\Delta]}{(1-m\hbar\omega/\Delta)[1+U_p/\Delta-(1-m)\hbar\omega/\Delta]}
\\
{\cal R}&=\sum_m
\frac{{\cal J}_m(\kappa)^2}
{1+U_p/\Delta-m\hbar\omega/\Delta}.
\end{align}
Because these processes can be Pauli blocked, they also generate an interaction between the $d$ fermions and the $P$ dimers. After the uniform Lamb shift has been absorbed into $\delta_R$, the remaining interaction term is
\begin{align}
H_{Pd}={}&
\frac{2 t^2}{\Delta}({\cal T}-{\cal R})
\sum_{\langle ij\rangle}P_{\langle ij\rangle}^\dagger P_{\langle ij\rangle}
n_{[i+j]}.
\label{eq:blockstructure}
\end{align}

Collecting these terms, the first elimination gives, to leading order in $t/\Delta$ and $U_d/\Delta$,
\begin{align}
H_{\rm md}={}&H_d+\delta_R\sum_{\langle ij\rangle}
P_{\langle ij\rangle}^\dagger P_{\langle ij\rangle}
\nonumber\\
&+\Lambda\sum_{\langle ij\rangle}
(P_{\langle ij\rangle}^\dagger D_{[i+j]}+\mathrm{h.c.})+H_{Pd},
\label{eq:md}
\end{align}
where $H_d$ is the square-lattice Hubbard Hamiltonian
\begin{align}
H_d={}&-t_{\rm eff}\sum_{\langle ij\rangle,\sigma}
(d^\dagger_{i\sigma}d_{j\sigma}+\mathrm{h.c.})
+U_d\sum_i n^d_{i\uparrow}n^d_{i\downarrow}.
\label{eq:Hd}
\end{align}
This is a two-channel lattice Feshbach model and must be retained whenever $|\delta_R|$ is comparable to the pair-conversion amplitude, the $d$-bandwidth, or the blocking interaction, which are all $\mathcal{O}(t^2/\Delta)$.

For a moderately large positive dressed detuning,
\begin{equation}
\Delta \gg \delta_R \gg t^2/\Delta,
\label{eq:stage2}
\end{equation}
the $P$-dimer states lie at higher energy and can be eliminated in a second-stage expansion, in which case $H_{Pd}$ is no longer important. This gives a $d$-only effective model on the vertex square lattice:
\begin{align}
H_{\square}&=H_d-2J\sum_{\langle ij\rangle}
D^\dagger_{[i+j]}D_{[i+j]}, \label{eq:target}
\end{align}
where $J=|\Lambda|^2/(2\delta_R)>0$. Nonresonant fourth-order corrections are $O(t^4/\Delta^3)$ and thus subleading in the assumed hierarchy of scales. The bond operator with strength $J$ contains an AF exchange
$+J\,\bm S_i\cdot\bm S_j$, density and correlated-hopping interactions, and
pair hopping (End Matter). Decomposed in the Cooper channel, it induces attraction in both the $d_{x^2-y^2}$-wave and $s$-wave channels. The Hubbard repulsion $U_d$ penalizes the on-site component of the $s$-wave channel and thereby exposes the $d$-wave pairing channel, allowing $U_d/J$ to tune between $s$ and
$d$ pairing.

The models in Eqs.~\eqref{eq:md} and \eqref{eq:target} are structurally identical to effective models descending from the Lieb--Holstein model~\cite{HanKivelson2023}. Reference~\cite{HanKivelson2023} also identifies a controlled route to a quantum dimer model, and hence to RVB physics, when the $P$ dimers lie below the $d$ manifold. This corresponds here to a negative detuning, $\delta_R<0$. Such a regime is in principle accessible in the present setup, but it requires preparing a distinct prethermal branch with a substantial initial population of $P$ dimers. The dimer-monomer model in Eq.~\ref{eq:md} is also similar to what was used to realize spin liquid states in a Bosonic Lieb-Hubbard setup~\cite{karch2026dynamicalpreparationu1quantum}.

\begin{figure}[t]
\centering
\includegraphics[width=0.98\linewidth]{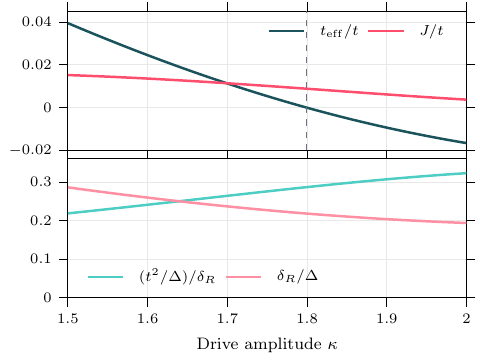}
\caption{\textbf{Interference control of the effective couplings.}
The upper panel shows the signed effective hopping $t_{\rm eff}/t$ and mediated interaction $J/t$ for $1.5\leq\kappa\leq2$. The lower panel shows the control ratios $(t^2/\Delta)/\delta_R$ and $\delta_R/\Delta$. The vertical dashed line in the upper panel marks the interference zero of $t_{\rm eff}$ at $\kappa\approx 1.8$. All photon harmonics with $|m|\leq60$ are retained for $t=1$, $\Delta=4$, $U_p=4$, and $\hbar\omega=10.75$. The displayed hierarchy ratios remain below $0.32$ throughout the plotted window.}
\label{fig:control}
\end{figure}

\textit{Controlling the coupling strength.---} We aim to simulate the intermediate-coupling regime of Eq.~\eqref{eq:target}, where competing orders can emerge at intermediate temperature scales. At first sight, the effective model seems incompatible with this goal. For a generic drive amplitude, $t_{\rm eff}$ and $\Lambda$ are both $O(t^2/\Delta)$, suggesting
\begin{align}
\frac{J}{|t_{\rm eff}|}
&=\frac{|\Lambda|}{2|t_{\rm eff}|}\frac{|\Lambda|}{\delta_R}
\stackrel{?}{\sim}\frac{|\Lambda|}{\delta_R} \ll 1
\end{align}
in the controlled second-elimination limit. The additional control provided by $\kappa$, however, allows $J/|t_{\rm eff}|\sim1$ while preserving the perturbative hierarchy. The key observation is that the all-harmonic sum for $t_{\rm eff}$  has interference zeros at which $\Lambda$ remains nonzero. Let $\kappa_*$ denote such a zero. Choosing $\kappa$ close, but not equal, to $\kappa_*$ can enhance $|\Lambda|/(2|t_{\rm eff}|)$ enough to compensate for the small ratio $|\Lambda|/\delta_R$. This provides access to the intermediate-coupling regime
\begin{align}
|t_{\rm eff}|\sim J\sim U_d.
\end{align}
Figure~\ref{fig:control} illustrates this lever using the complete harmonic sums: the ratio $J/|t_\text{eff}|$ can indeed be tuned by $\kappa$ in a wide range while keeping the control parameters small. A caveat, though, is that this tuning also suppresses the overall energy scale of the effective model. An experiment must therefore choose a finite compromise between interaction-to-hopping ratio, absolute temperature scale, and lifetime.

\begin{figure}[t]
\centering
\includegraphics[width=0.98\linewidth]{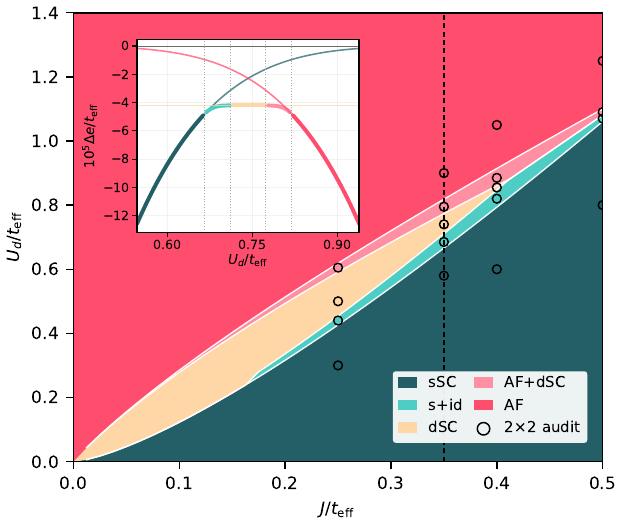}
\caption{\textbf{Zero-temperature mean-field phase diagram of \eqref{eq:target}.}
Lowest-energy variational state at half filling over the displayed coupling window. Distinct colors denote $s$SC, time-reversal-breaking $s+id$, $d$SC, AF+$d$SC, and AF.   (Inset) Condensation energy relative to the symmetric normal saddle along the dashed $J/t_{\rm eff}=0.35$ cut. Thin curves are stationary solutions and thick segments are selected minima; vertical dotted lines mark refined phase boundaries.}
\label{fig:mf}
\end{figure}

\textit{Mean-field calculation.---} To test the tendencies toward $d$SC and AF in the square-lattice model, Eq.~\eqref{eq:target}, we solve the saddle point equations for multiple orders at zero temperature, and audit the results with unbiased Gaussian states using a repeated $2\times2$ supercell. All particle--hole and particle--particle bilinears allowed within this cell, together with their arbitrary coexistence, are treated on an equal footing. The End Matter gives further details of the calculation. Figure~\ref{fig:mf} summarizes the resulting phase diagram, which contains regimes of $s$SC, $d$SC, AF, and coexistence.

These mean-field results are not controlled at intermediate coupling and should therefore be regarded as suggestive. It is nevertheless notable that the phase diagram is qualitatively consistent with beyond-mean-field studies of closely related models descending from electron--phonon problems. Assaad's quantum Monte Carlo study of a bond-hopping-square interaction found a half-filled evolution from $d$SC to AF as the repulsion increases~\cite{Assaad1998}. Sign-problem-free simulations show that bond SSH phonons generate AF exchange and order~\cite{CaiLiYao2021}, while DMRG finds robust $d$SC upon doping from the combined SSH coupling and Hubbard repulsion, between AF and $s$SC regimes~\cite{WangJiangYao2025}. This agreement across related models and independent methods makes the competing tendencies plausible, while leaving their precise boundaries and transition structure as strongly correlated many-body questions for the Floquet simulator.

\textit{Preparation and limitations.---} Quasienergies are periodic modulo $\hbar\omega$, so an exact Floquet spectrum has no canonical ground state~\cite{Hone1997}.  Here ``ground state'' means the low-entropy state of the branch-selected prethermal Hamiltonian. At zero drive the conversion $\Lambda$ vanishes, while the positive charge-transfer gap makes the cooled gas $d$-orbital dominated.  For $\delta_R>0$---a red-detuned drive---this initial state connects continuously to the lower, $d$-like branch of the resonant pair-dimer block.  The modulation should be ramped while the dressed mediator energy remains positive, with the frequency adjusted if necessary to compensate its amplitude-dependent Lamb shift.  The observation window must satisfy
\begin{equation}
\tau_{\rm ramp},\ \hbar/J\ll\tau_{\rm heat}.
\label{eq:lifetime}
\end{equation}

Several experimental risks are independent and should not be hidden by the formal scaling limit. First, unwanted one-particle, separated-pair, and many-body resonances must remain outside the effective bandwidth. Second, the required modulation amplitude can couple to higher Wannier bands even when the tight-binding derivation remains controlled. Third, real $p$-doublon population, spatially varying Lamb shifts, and interaction-dependent detuning can broaden the resonance. All these factors require a careful assessment for a specific experimental implementation.

\textit{Discussion and outlook.---} Our work establishes a controlled route from repulsive microscopic interactions to a structured attraction with flexibly tunable strength. This setup may expose the competition between AF and $d$SC directly in a quantum simulator. Our approach is complementary to the route studied in Ref.~\cite{Herre2026}, where circular light driving modifies photoassisted hopping, the Fermi surface, and the magnetic fluctuation spectrum; within Floquet random-phase approximation (RPA), $d+id$ pairing appears near the AF boundary and evolves toward triplet $p$-wave pairing near a Floquet Lifshitz transition. The drive there acts primarily through the one-particle dispersion, so the pairing kernel is inherited indirectly from the driven magnetic fluctuations and is tied to the accompanying band reconstruction. Our construction instead focuses on engineering the effective interactions. Related works also explored schemes suppressing hopping relative to superexchange in the large-\(U\) limit \cite{Coulthard2017}; near half filling, however, this hierarchy freezes charge dynamics and may not lead to $d$SC~\cite{EmeryKivelsonLin1990,HanKivelsonYao2020}.

Conceptually, our work also develops an intimate connection between Floquet and electron--boson problems. We expect this bridge to provide further design principles for driven quantum matter. For example, the same kind of bond electron-phonon models have been argued to host high-temperature SC~\cite{Sous2018,ZhangSous2023,KimHanSous2024,WangJiangYao2025} away from half-filling. This physics may be studied in the current setup in an analog way.

{\bf Acknowledgements. } We thank Clemens Kuhlenkamp and Carl Zelle for helpful discussions. This research was supported by the U.S. National Science Foundation grant No. DMR 2245246. Z.~H. was supported by the Gordon and Betty Moore Foundation EPiQS Award 8683. OpenAI Codex (GPT-5.6) was used interactively to assist with algebraic checks, the development and execution of mean-field and plotting code, and language revision. The authors reviewed and revised all AI-assisted material.

\makeatletter
\def\bibfont{\footnotesize\@clubpenalty\clubpenalty}
\makeatother
\bibliography{references}

\clearpage
\appendix
\setcounter{equation}{0}
\renewcommand{\theequation}{E\arabic{equation}}

\section{End Matter}
\label{sec:endmatter}
\subsection{Details of the virtual processes}

In rotating frame formulation of the problem, provided that none of the appreciably weighted singly occupied \(p\) states is resonant, this sector can be eliminated perturbatively. The resulting second-order processes generate the effective hopping \(t_{\rm eff}\), the pair-conversion amplitude \(\Lambda\), and the diagonal shifts entering \(\delta_R\) and \(H_{Pd}\), as given in the main text. The effective hopping arises when a fermion initially on a \(d\) orbital virtually enters the intervening \(p\) orbital and subsequently exits onto the neighboring \(d\) orbital. Because the initial and final states belong to the same Floquet replica, the two hops carry the harmonic weight \({\cal J}_m(\kappa)^2\) and the intermediate singlon has energy denominator \(\Delta-m\hbar\omega\). Pair conversion instead involves two fermions from the symmetric \(d\)-bond orbital successively entering the same \(p\) orbital. The two hops together transfer one drive quantum, which can be partitioned between them as \(m\) and \(1-m\); the corresponding paths therefore interfere with weights \({\cal J}_m(\kappa){\cal J}_{1-m}(\kappa)\). Since the initial \(d\)-bond pair and final \(P\) dimer are not exactly degenerate, the Hermitian second-order matrix element contains denominators referenced to both endpoint energies, as retained in the expression for \(\Lambda\) in the main text. The enhanced hybridization of the normalized symmetric \(d\)-bond orbital and the two spin orderings account for its overall prefactor. The remaining second-order processes are diagonal. When the bond-center \(p\) orbital is empty, a fermion can execute the sequence \(d\rightarrow p\rightarrow d\), producing the ordinary \(d\)-sector renormalization and the shift of a \(d\)-bond pair. When the \(p\) orbital contains a \(P\) dimer, the allowed sequence is instead \(p\rightarrow d\rightarrow p\), which produces the dimer Lamb shift. The difference between these shifts renormalizes the pair-transfer detuning from \(\delta\) to \(\delta_R\). Their dependence on the occupations of both orbitals also leaves the operator-valued blocking term \(H_{Pd}\): a \(P\) dimer removes the \(d\rightarrow p\rightarrow d\) path, while occupation of \(d\) Pauli-blocks the reverse \(p\rightarrow d\rightarrow p\) process. 

\subsection{Effective interaction content}

The compact bond attraction can be decomposed as
\begin{align}
&-2J\sum_{\langle ij\rangle}
D_{[i+j]}^\dagger D_{[i+j]}\nonumber\\
={}&-2J\sum_i n_{i\uparrow}n_{i\downarrow}
-\frac{J}{4}\sum_{\langle ij\rangle}n_i n_j +J\sum_{\langle ij\rangle}\bm S_i\cdot\bm S_j 
\nonumber\\
&-\frac{J}{2}
\sum_{\langle ij\rangle,\sigma}
B_{ij,\sigma}(n_{i\bar\sigma}+n_{j\bar\sigma})-\frac{J}{2}\sum_{\langle ij\rangle}
\left(
d^\dagger_{i\uparrow}d^\dagger_{i\downarrow}
d_{j\downarrow}d_{j\uparrow}
+\mathrm{h.c.}
\right).
\label{eq:EMdecomp}
\end{align}
where $
B_{ij,\sigma}
=d^\dagger_{i\sigma}d_{j\sigma}+\mathrm{h.c.}$
is the hopping operator. This coupling therefore generates an antiferromagnetic exchange, reduces the net local repulsion to $U_d-2J$, and produces nearest-neighbor density, correlated-hopping, and pair-hopping interactions. 

At zero center-of-mass momentum, the pairing kernel generated by the compact
interaction is
\begin{equation}
V(\bm k,\bm k')
=-2J\sum_{\alpha=x,y}
(1+\cos k_\alpha)(1+\cos k'_\alpha).
\label{eq:EMpairkernel}
\end{equation}
Introducing the $A_{1g}$ and $B_{1g}$ form factors
\begin{equation}
f_s(\bm k)=2+\cos k_x+\cos k_y,
\qquad
f_d(\bm k)=\cos k_x-\cos k_y,
\label{eq:EMformfactors}
\end{equation}
the kernel separates exactly as
\begin{equation}
V(\bm k,\bm k')
=-Jf_s(\bm k)f_s(\bm k')
-Jf_d(\bm k)f_d(\bm k').
\label{eq:EMchanneldecomp}
\end{equation}
The first term belongs to the fully symmetric $A_{1g}$ sector and contains
both on-site and extended-$s$ components. The second is a pure $B_{1g}$ or $d_{x^2-y^2}$ channel.

\subsection{Numerical method}

We determine the zero-temperature phase diagram of Eq.~\eqref{eq:target} at half filling, setting \(t_{\rm eff}=1\). The quantitative phase boundaries are obtained from a symmetry-adapted Gaussian calculation that retains the five candidate saddles found in the scan: \(s\)SC, \(s+id\), \(d\)SC, AF+\(d\)SC, and AF. For each \(J/t_{\rm eff}=0,0.0125,\ldots,0.5\), we construct the corresponding Bogoliubov--de Gennes Hamiltonian from the normal bond expectation values, staggered magnetization, on-site anomalous average, and complex nearest-neighbor singlet amplitudes. The resulting fixed-point equations are solved to determine the ground state. We use logarithmic variables for small order parameters, continuation from neighboring parameter points, and multiple initial seeds. The lowest energy per site among the converged saddles is selected.

The boundaries are refined directly rather than read from a finite grid in \(U_d\). The \(s\)SC--\(s+id\) and \(s+id\)--\(d\)SC boundaries are located where the linearized subdominant pairing eigenvalue in the corresponding self-consistent superconducting background reaches unity. Similarly, the two edges of the AF+\(d\)SC region are obtained from the AF instability of the \(d\)SC saddle and the \(d\)-wave instability of the AF saddle. If two locally stable saddles compete before either loses stability, we instead locate the zero of their energy difference. At small \(J\), the superconducting gaps and the \(s+id\) coexistence window become exponentially small; when that window cannot be resolved nonlinearly, we report the crossing of the linearized \(s\)- and \(d\)-wave gap scales and set its displayed width to zero. The Brillouin-zone integrals use an infrared-adapted quadrature with logarithmically spaced panels resolving both the Fermi surface and the half-filled van Hove points. Increasing the outer quadrature from 32 to 64 nodes and the infrared resolution from 14 to 18 decades shifts the reported boundaries by less than \(1.5\times10^{-3}t_{\rm eff}\). As a separate benchmark, the \(J=0\) calculation recovers the nested Hubbard result: every \(U_d>0\) has an AF saddle below the normal state.

We then audit, rather than determine, these boundaries with an unrestricted periodically repeated \(2\times2\) supercell. For each of the four site orbitals and eight positively oriented nearest-neighbor bonding orbitals, this calculation retains the complete spin-resolved \(2\times2\) normal density matrix and the complex same-orbital singlet anomalous density. These are all normal and anomalous self-energies generated by the Gaussian variational derivative of the original compact interaction, without imposing phase labels before convergence. Together with the chemical potential, the stationary problem contains 73 real variables and a \(16\times16\) Bogoliubov--de Gennes Hamiltonian. We first performed a coarse unrestricted scan on a \(6\times16\) grid covering \(0\leq J/t_{\rm eff}\leq0.5\) and \(0\leq U_d/t_{\rm eff}\leq1.5\). We then carried out 114 higher-resolution optimizations at the 19 parameter points marked by open circles in Fig.~\ref{fig:mf}, using six starts per point: the symmetry-adapted solution and charge-density-wave, valence-bond, nematic, pair-density-wave, and random mixed textures. All runs converged with residuals below \(2.0\times10^{-7}\); no lower-energy translation-breaking solution was found, and independently converged energies agreed within \(4.2\times10^{-9}t_{\rm eff}\) per site. These audits substantially reduce initialization and supercell bias but do not exclude disconnected minima, incommensurate order, periods longer than \(2\times2\), or phase separation. 

\subsection{Branch selection and operational ground state}

The local pair-conversion problem is described by
\begin{equation}
H_{\rm loc}
=
\begin{pmatrix}
0&\Lambda^*\\
\Lambda&\delta_R
\end{pmatrix},
\qquad
E_\pm
=\frac{\delta_R\pm\sqrt{\delta_R^2+4|\Lambda|^2}}{2}.
\label{eq:EMbranches}
\end{equation}
For $\delta_R>0$, the lower eigenstate $E_-$ is predominantly a $d$-bond
pair and is shifted downward by $-|\Lambda|^2/\delta_R$. This produces the
attractive interaction and corresponds to a red-detuned drive. For
$\delta_R<0$, the $P$ dimer instead forms the lower branch; the $d$-like
state is $E_+$ and is shifted upward by $+|\Lambda|^2/|\delta_R|$. Shifting
either quasienergy by an integer multiple of $\hbar\omega$ changes its
representative in the Floquet zone but not its orbital composition or its
connection to the initial state.

With the modulation switched off, $\Lambda=0$, and the large positive
charge-transfer gap makes the cooled system predominantly $d$-orbital in
character. The proposed protocol ramps the drive from this state while
keeping $\delta_R>0$, so that the prepared state remains connected to the
lower, $d$-like branch. For a ramp dominated by the variation of $\Lambda$
at approximately fixed $\delta_R$, the local adiabaticity and prethermal
conditions are
\begin{equation}
\frac{\hbar|\dot{\Lambda}|}{\delta_R^2}\ll1,
\qquad
\tau_{\rm ramp}\ll\tau_{\rm heat}.
\label{eq:EMramp}
\end{equation}
One must additionally impose adiabaticity relative to any many-body gap
possessed by the target state. In a gapless thermodynamic phase, the
operational objective is instead to preserve a sufficiently low entropy
during the ramp. Floquet--Gibbs cooling is not generic because an external
bath may exchange both energy and drive quanta with the system
\cite{Shirai2015,Seetharam2015}. The relevant notion of a ground state is
therefore the low-entropy state of a branch-selected prethermal Hamiltonian,
not the state of lowest exact quasienergy.

\subsection{Validity conditions}

In addition to $t/\Delta\ll1$, the construction implicitly requires the following
conditions:
\begin{enumerate}
\item For every appreciably weighted harmonic, the singlon denominators
$|\Delta-m\hbar\omega|$ and
$|\Delta+U_p-m\hbar\omega|$ must remain large compared with the corresponding
Bessel-weighted tunneling matrix elements and the relevant low-energy
bandwidths.
\item The separated-particle detuning
$|2\Delta-\hbar\omega|=|U_p-\delta|$ must exceed the effective two-particle
bandwidth, so that the drive selectively addresses the on-orbital $P$ dimer
rather than two spatially separated $p$ fermions.
\item The retained energy scales must satisfy
$U_d,|t_{\rm eff}|,J\ll\hbar\omega$, and no combination resonance involving
$U_d$ may fall inside the prethermal bandwidth.
\item Both the modulation amplitude $A=\kappa\hbar\omega$ and the drive
frequency must remain below the relevant gaps to unwanted Wannier bands.
\item The heating time must satisfy
$\tau_{\rm heat}\gg\tau_{\rm ramp},\hbar/J$.
\end{enumerate}
The first three conditions isolate the desired Sambe crossing and justify
the effective low-energy description. The fourth protects the assumed
orbital Hilbert space, while the fifth is a dynamical requirement rather
than a property of the Schrieffer--Wolff expansion. Satisfying all these
conditions while retaining a useful absolute value of $J$ is the central
practical challenge of the proposal.

\end{document}